\documentclass{ctic}

\usepackage{graphicx}

\title{Digital Curvatures Applied to 3D Object Analysis and Recognition: A Case Study}

\author{Li Chen$^{1}$, Soma Biswas$^{2}$}

\address{$^{1}$~University of the District of Columbia.\\
$^{2}$~University of Maryland.\\
\texttt{lchen@udc.edu} ,\texttt{somabapi@gmail.com} 
}

\usepackage{graphics} 
\usepackage{epsfig} 

\newtheorem{theorem}{Theorem}[section]

\newtheorem{lemma}[theorem]{Lemma}

\begin{document}

\maketitle
\thispagestyle{empty}

\begin{abstract}
In this paper, we propose using curvatures in digital space for 
3D object analysis and recognition. Since direct adjacency has only  
six types of digital surface points in local configurations, 
it is easy to determine and classify the discrete curvatures
for every point on the boundary of a 3D object. 
Unlike the boundary simplicial decomposition (triangulation), the curvature can take any real value.
It sometimes makes difficulties to find a right value for threshold. 
This paper focuses on the global properties of categorizing curvatures for small regions.
We use both digital Gaussian curvatures and digital mean curvatures to 3D shapes.
This paper proposes a multi-scale method for 3D object analysis and a vector method for 3D
similarity classification. We use these methods for face recognition and 
shape classification. 
We have found that the Gaussian curvatures mainly describe the global features and average characteristics
such as the five regions of a human face. However, mean curvatures
can be used to find local features and extreme points such as nose in 3D facial data.

\end{abstract}


\section{Introduction}

In recent years, there are many applications related to geometric analysis in 3D image processing, especially
the methods using topological and geometric invariants. \cite{Bis} \cite{Hil}
 This paper presents a preliminary study of using digital curvatures for digital surfaces 
to classify different 3D objects. 

A 3D object can be represented by one or several closed surfaces (2D-manifolds). Curvatures
that describe the degree of change at a point on the surface have been used for many years in 
3D image processing~\cite{Bes} \cite{Del95}  \cite{Mok}.  
The typical technology related to curvatures is the following: (1) to use triangulation of the surface, 
(2)to fit the digital image to a continuous surface (using B-spline), and (3) to calculate   
the standard Gaussian and/or Mean Curvatures. 
This usually refers to the areas of computational geometry.

On the other hand, methods in digital and discrete geometry have been developed rapidly in recent years
for 3D computer graphics and
3D image processing~\cite{KR}. For example, recognition algorithms related to 3D 
medical images can be found in ~\cite{BK08} in which
Brimkov and Klette made extensive investigations in boundary tracking.

In this paper, we propose using curvatures in digital space for  3D object matching, classification, and recognition. 
Since direct adjacency only has six types of digital surface points in local configurations, 
it is much easier to determine and classify the digital curvatures for every point on the boundary surface of 
a 3D object. Because of the limited number of local configurations, this method may only apply to certain  applications.

\section{Background}

Given a set of cloud points in 3D, we assume that they are connected. (Such is the case for 
3D MRI brain images.)  Since  cubical space with direct adjacency, or (6,26)-connectivity space, has the simplest 
topology in 3D digital spaces, we will use it as the 3D image domain. 
It is also believed to be sufficient for the topological 
property extraction of digital objects in 3D.  In this space, two points are said to be adjacent in 
(6,26)-connectivity space if the 
Euclidean distance between these two points is 1.
 
Let $M$ be a closed (orientable) digital surface in the 3D grid space in direct adjacency. 
We know that there are exactly 6-types of digital surface
points~\cite{Che04}\cite{CR}.

\begin{figure}[h]
	\begin{center}

   \epsfxsize=2in 
   \epsfbox{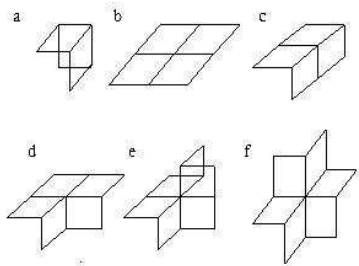}

	\end{center}
\caption{ Six types of digital surfaces points in 3D}
\end{figure}

Assume that $M_i$ ($M_3$, $M_4$, $M_5$, $M_6$) is the set of 
digital points with $i$ 
neighbors. We have the following result for a simply connected 
$M$ ~\cite{Che04}:

\begin{equation}
          |M_3| =8 + |M_5| + 2 |M_6| .      
\end{equation}

 \subsection{ Gaussian Curvatures and Mean Curvatures}

If $K_i$ denotes the Gaussian curvature of elements in $M_i$, $i=$ 3,4,5,6. We have
(see discussion in \cite{CR}.)

\begin{lemma}\label{l21}
   (a) $K_3  = \pi/2$,
  (b) $K_4= 0$,  for both types of digital surface points,
  (c) $K_5  = - \pi /2$, and
   (d) $K_6  = - \pi$,  for both types of digital surface points.
\end{lemma}

\noindent We also have a genus formula based on the Gauss-Bonnet Theorem \cite{CR}

\begin{equation}
           g = 1+ (|M_5|+2 \cdot |M_6| -|M_3|)/8. 
\end{equation} 

If $H_i$ denotes the mean curvature of elements in $M_i$, $i= 3,4,5,6.$ We have

\begin{lemma}\label{l21}
   (a) $H_3  =  \frac{4}{\sqrt{3}}$.
   (b) $H_4= 0$ for flat neighborhood, and $H_4= \sqrt{2}$ for a bended neighborhood.
   (c) $H_5  = \frac{4}{5}$ .
   (d) $H_6  = 0$,  for both types of digital surface points.
\end{lemma}

This lemma can be proved easily based on the formula derived by Meyer et al in 2002~\cite{May}.  
Goodman-Strauss and Sullivan  used the $H_6=0$ points to construct cubic minimum surfaces~\cite{GS}. 
A minimum surface can be defined as a surface
whose mean curvature at every point is 0. Here we give a brief proof for Lemma 2.2.

The formula derived in \cite{May} for the mean curvature normal at point $x$ is: 

\begin{equation}
  H(x)\cdot {\bf n} = \frac{1} { 2\cdot A} \Sigma_{x_i \in N(x)} (\cot \alpha_i + \cot \beta_i) (x-x_i) 
\end{equation}

\noindent where $N$ is the set of (discrete) neighbors of $x$.  $\alpha_i$ and $\beta_i$ are angles 
in two different triangles that are both opposite to line segment $x x_i$; 
which is  shared by the triangles.  $A$ is called the voronoi region of $x$. 
For digital space, the voronoi region is easy to determine. See Fig 2.  

\begin{figure}[h]
	\begin{center}

   \epsfxsize=2in 
   \epsfbox{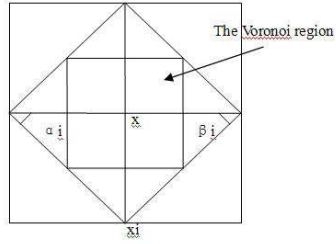}

	\end{center}
\caption{ The voronoi region in digital space}
\end{figure}

\begin{equation}
  H(x)\cdot {\bf n} = \frac{1} { A} \Sigma_{x_i \in N(x)} (x-x_i) 
\end{equation}

Note that $H(x)\cdot {\bf n}$ is a vector.  $|(x-x_i)|=1$ and $\cot \alpha_i$ = $\cot \beta_i$ =1. in Fig 2.
$A= \frac{1}{4} \cdot i$ for $M_i$ points.   $|\Sigma {x_i \in N(x)} (x-x_i)|=\sqrt(3)$ for $M_3$;  
$|\Sigma {x_i \in N(x)} (x-x_i)|=0$ for $M_4$ flat, and $\sqrt(2)$ for $M_4$ bended; 
$|\Sigma {x_i \in N(x)} (x-x_i)|=1$ for $M_5$;$|\Sigma {x_i \in N(x)} (x-x_i)|=0$ for $M_6$;
Therefore, $H_3  =  \frac{1}{3\cdot \frac{1}{4}} \sqrt(3)=\frac{4}{\sqrt(3)}$.
For bended $M_4$ points (Fig. 1 (c)), $H_4  =  \frac{1}{4 \frac{1}{4}} \sqrt(2)=\sqrt(2)$.
$H_5  =  \frac{1}{5 \frac{1}{4}}\cdot 1=\frac{4}{5}$.

\subsection{Digital Principal Curvatures}

The principal curvatures at a point $p$ of a surface, denoted $k_1$ and $k_2$, are the maximum and minimum values 
of the curve curvature on normal planes (intersecting with the surface). The relationship among the 
principal curvature, the Gaussian curvature, and the mean curvature are: $K=k_1\cdot k_2$ and $H=(k_1+k_2)/2$.
Therefore \cite{Kre}\cite{May}, 

\[ k_1 = H+\sqrt{H^2-K},   k_2 = H-\sqrt{H^2-K} \]

It is easy to get,

\begin{lemma}\label{l21}
   (a) $k_1^{(3)}  =  \frac{4}{\sqrt{3}} + \sqrt { \frac{16}{3} - \frac{\pi}{2}} = 4.24913$, 
       $k_2^{(3)}  =  \frac{4}{\sqrt{3}} - \sqrt { \frac{16}{3} - \frac{\pi}{2}} = 0.369675$.
   (b) $k_1^{(4b)}  = 0$, 
       $k_2^{(4b)}  = 0$;
       $k_1^{(4c)}  = 2\sqrt{2} = 2.82843$, 
       $k_2^{(4c)}  = 0$. 
   (c) $k_1^{(5)}  = \frac{4}{5} + \sqrt{ \frac{16}{25} +\frac{\pi}{2} }=2.28687 $, 
       $k_2^{(5)}  = \frac{4}{5} - \sqrt{ \frac{16}{25} +\frac{\pi}{2} }=-0.686875 $,  .
   (d) $k_1^{(6)}  =  \sqrt{  \pi } = 1.77245$, 
       $k_2^{(6)}  = -\sqrt{  \pi } =-1.77245$ for both types of $M_6$ digital surface points.
\end{lemma}

From these lemmas, we can find some interesting information about  digital 
principal curvatures.  $k_1^{4c}$ does not have $\pi$ involved in the fomula. However, from
Fig. 1c to Fig. 1a is just like to make a 90 degree angle similar to the case from Fig. 1b to Fig. 1c. 
Where the $\pi$ came from? Due to the fact of digital$\/$discrete mean curvatures were obtained from
an approximation, we think that the value of the mean curvatures for $M_{i}$ points can be modified as
$H'=c_1\cdot H +c_2$ as long as we keep $H'_3>H'_{4c}>H'_5>H_6=0$, where $c_1$ and $c_2$ are two constants.

\section{ Digital Curvatures and 3D Image Analysis} 

Our research is to investigate whether the total number of $M_i$ points mean anything to
the object. We have discussed Gaussian curvature in the last section. For mean curvatures, 
$M_i$ also contains information about mean curvatures. For instance in 
a closed digital surface, $M_4$ is independent to $M_3$; $M_4$ contains parabolic points (the bended ones).
$M_3$ contains the elliptical points since the surface point is locally convex. 
$M_5$ and $M_6$ contain hyperbolic points; 
the Gaussian curvature is negative and the surface point will be locally saddle shaped.

\subsection{Analysis Based on Digital Gaussian Curvatures}
 
Let $S$ be a subset of the 2D manifold $M$. We define a feature vector called the digital curvature vector
 $f_{S}= (m_3,m_4,m_5,m_6)$, where $m_i =  |M_i|$ with respect to $S$. We can also split
 the vector into a six component vector to include two cases for $M4$ and $M6$ respectively.
Basically, the vector now contains four components: $f_{S}=(|M_3|,|M_4|,|M_5|,|M_6|)$.

The motivation to define such a vector for a local region is based on the corresponding 
Gaussian-Bonnet theorem \cite{Kre}. Suppose $C$ is a boundary
curve of $S$ that is simply connected and $k_g$ is the geodesic curvature of $C$, then

\begin{equation}
    \int_{C} k_{g} dt + \int\int_{S} K d A= 2 \pi   
\end{equation}

If $C$ is a $n$-polygon and $\alpha_{i}$ is the interior angle, then 

\begin{equation}
       (n-2)\cdot \pi + \int_{C} k_{g} dt + \int\int_{S} K d A= \Sigma_{i=1}^{n} \alpha_{i} 
\end{equation}

\noindent Therefore,
\begin{equation}
  g_S = g(f_{S})=|M_3|\cdot K_3 + |M_5|\cdot K_5+|M_6|\cdot K_6) = \int\int_{S} K d A
\end{equation}

\noindent can also be used to represent the total geodesic curvature of $C$ : 
\begin{equation}
  \int_{C} k_{g} dt = 2 \pi - g_S  
\end{equation}

In image processing, we can define $S$ (or $C$) as a rectangular region. If $S$ only contains
one surface point, then $g_S$ is just $K$.     
In practice, the vector $f_{S}$ can be used for a closed surface or a small region centered 
at one point. Between two regions, there can be intersections or no intersection. The region 
can be 2D or 3D, depending on whether the problem can be projected into 2D easily.  The 
most popular shapes of the domain regions are circles/spheres and rectangles/cubes.

We have used this technique to analyze human facial data.  
If the size of the region reduced by $1$, $2$, $2^2$, ..., $2^k$ times, we will get a sequence of $f_S$, or simple get $g_S$. 
Then we can see the change of the curvatures. Such a method is
usually called a multi-scaling method.

Let us look at the following example. The two original images are shown in Fig 3.

\begin{figure}[h]
	\begin{center}

   \epsfxsize=2in 
   \epsfbox{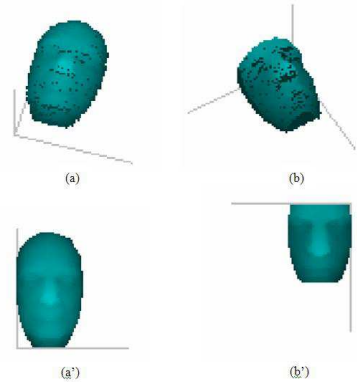}

	\end{center}
\caption{ Two original human faces from NIST-FRGC data sets}
\end{figure}

Fig 4 and Fig 5 show the initial curvature calculation by projection from
$x-$ and $y-$ directions.  $'*'=\pi/2$ indicates $M_3$-points; $'+'=-\pi/2$ indicates $M_5$-points;
$'X'=-\pi$ indicates $M_6$-points;

\begin{figure}[h]
	\begin{center}

   \epsfxsize=2in 
   \epsfbox{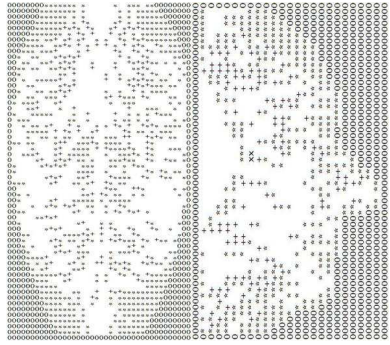}

	\end{center}
\caption{ Digital curvature on face-one's surface by projection }
\end{figure}

\begin{figure}[h]
	\begin{center}

   \epsfxsize=2in 
   \epsfbox{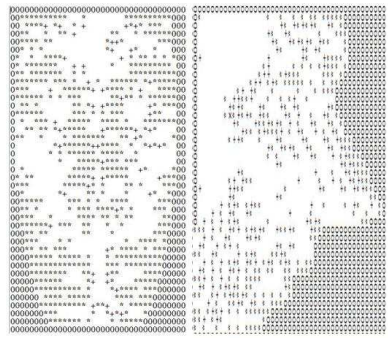}

	\end{center}
\caption{ Digital curvature on face-two's surface by projection }
\end{figure}

Fig 6. shows the projected Gaussian curvature data for each scale for face-one from $64\times 64\times 64$ 
to $8\times 8\times 8$. Fig 7 is for face-two. In order to be able to observe easily, we 
will display the same size image when the scale changes.

\begin{figure}[h]
	\begin{center}

   \epsfxsize=4in 
   \epsfbox{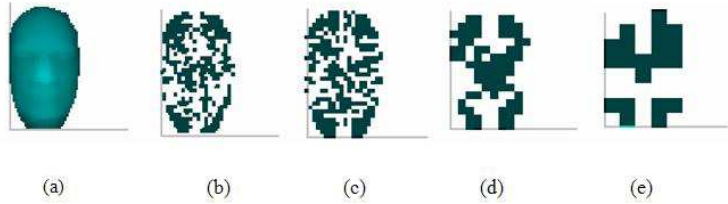}

	\end{center}
\caption{ Digital curvature scaling for Face-one }
\end{figure}

\begin{figure}[h]
	\begin{center}

   \epsfxsize=4in 
   \epsfbox{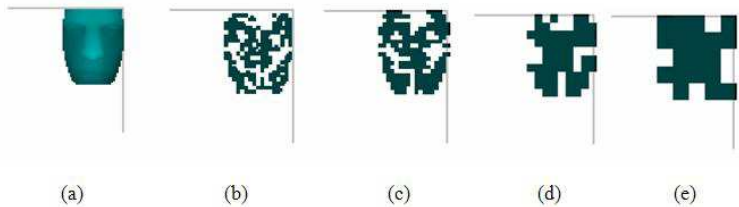}

	\end{center}
\caption{ Digital curvature scaling for Face-two }
\end{figure}

Analyzing Fig 7 and 8, we saw that the multi-scaling method can identify some of the interesting areas of the face. 
For example, we can identify five areas of interest for face-one and face-two.
They both contain two areas in the top and in the bottom and one area in the middle. 
These five areas indicate two region on the head-sides, two cheeks, and the nose. 
The nonzero total of Gaussian curvature for a region that remains
means that total change has not been canceled in this region. 

The result shows that the first person has a flatter face than the second person 
since the method reaches the five interesting areas earlier.
 We can also identify the flat-or-bending regions in the 
human face as shown in Fig. 9. A total of four such areas can be easily found.

\begin{figure}[h]
	\begin{center}

   \epsfxsize=3in 
   \epsfbox{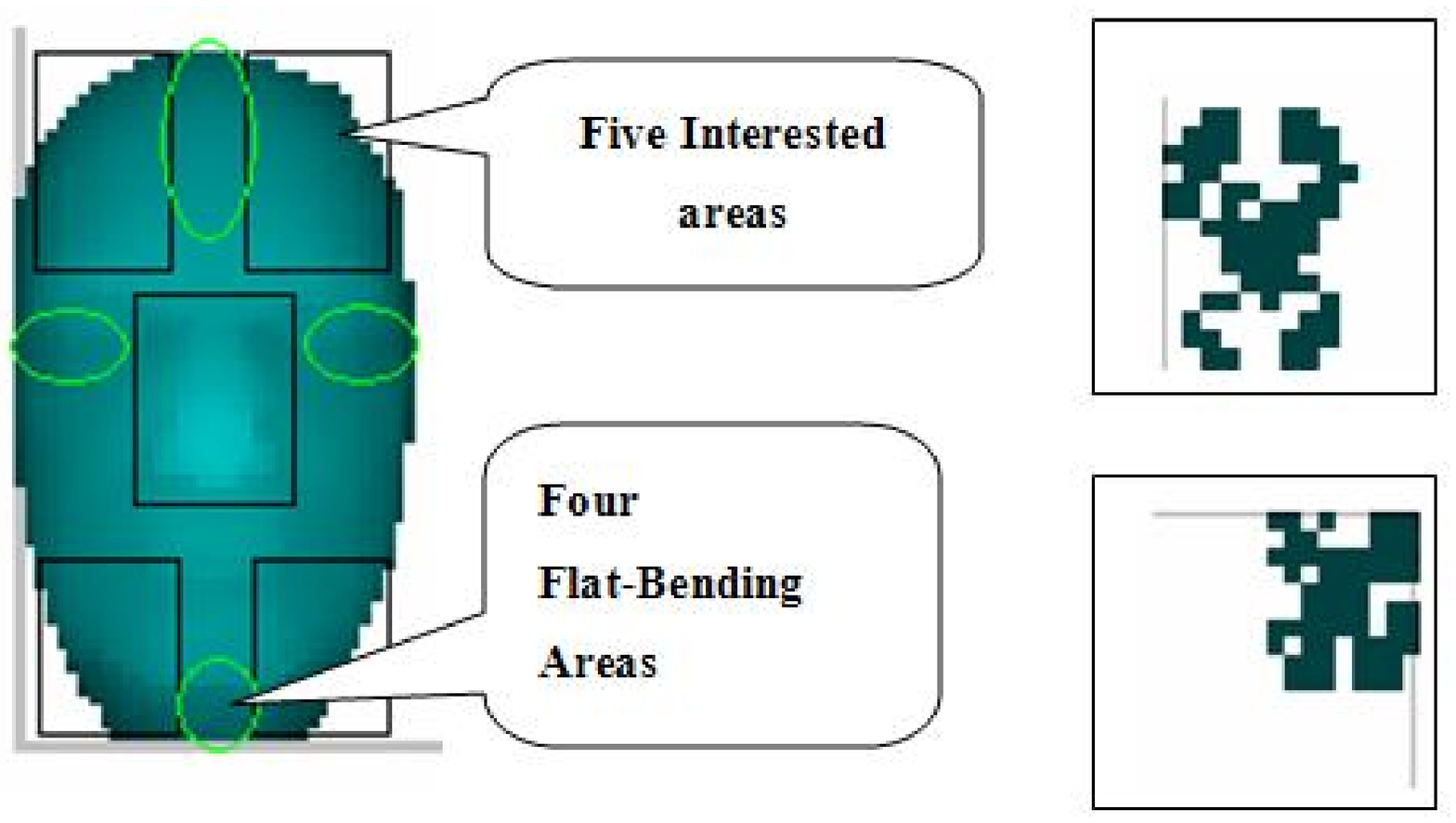}

	\end{center}
\caption{ Five areas of interests  and four flat-or-bending areas are 
          the common similar characteristics of human face}
\end{figure}

 The advantage of Gaussian curvature-based calculation is that it is not a simple processing of pixel averages in the region.
The curvature-based calculation is based on geometric and topological properties of the 3D object. For instance
we know the total Gaussian curvature will be a constant as the selected region becomes the whole 3D data
array. In summary, the method described in this section can identify the five regions of interest  
and four flat-or-bending regions as the common similar characteristics of a human face.
In addition, the calculation is much easier than that of triangulation based images.   The following lemma will
provide evidence for this. 
    
\begin{lemma}\label{l21}
 (1) The algorithm designed to obtain scaled local Gaussian curvatures for 
finding five areas of interest on the image of a human face is in $O(n log n)$. 
 The space needed is $O(log n)$. (2) There is also an algorithm to solve the same
problem in linear space and linear time. 
\end{lemma}

{\it Proof: } 
(1) Since the $2^k-1$-scaled Gaussian curvature data can be used 
to calculate the $2^k-1$-scaled Gaussian curvature data. We can design
a fast linear algorithm due to the reduction of the size of the array by half. 
In such a case, the space needed will also be $O(n)$ if we just use the input array and not the 
intermediate data. By using the output of the result for each resolution or scale, the time complexity will be
$O(n\log n)$ and the space complexity will be $O(\log n)$ (consider only  the space needed to run the algorithm). 
(2) If we keep all intermediate data, the space complexity will be enlarged to $O(n)$, but the time complexity
will be reduced to $2\cdot n$, where $n$ is the total number of points in the array. 
 
\subsection{Investigation Based on Digital Mean Curvatures}

For the mean digital curvature application, we have investigated the calculation based on the absolute 
average for small regions.  The geometric meaning of such treatment for digital mean curvature 
is the zigzagged points on the surfaces. 
Mean curvature zero points indicate the critical points that change from inward to outward if it is not a flat point. 
The result of an example is shown in Fig. 9.    

\begin{figure}[h]
	\begin{center}

   \epsfxsize=3in 
   \epsfbox{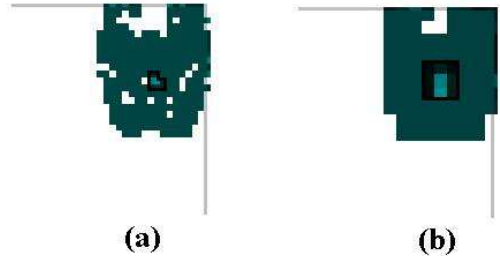}

	\end{center}
\caption{ Digital mean curvature to find a peak point that indicates nose of human face: 
          (a) Image is made by a $2\times 2$ summation of the absolute values of digital mean curvatures, 
          and (b)Image is made by a $4\times 4$ summation of the absolute values of digital mean curvatures.   }
\end{figure}

To summarize the findings, Gaussian curvatures describe the global features and the average characteristics
such as the five regions of a human face, but mean curvatures
find local features and extreme points such as the nose.

In the future, we will use this method for analyzing more human facial data. We can normalize the 
image size and calculate the distance for a pair of faces in corresponding scale or nearby scales. 
The following section will present a method for 3D data "rough" classification using digital curvature vectors.

\section{3D Object Classification Using Digital Curvature Vectors}

In this section, we will explore a method based on the digital curvature vectors 
for 3D shape similarity analysis and classification of 3D objects. 

There have been many research investigations that use local curvatures in 
3D shape similarity analysis.  Shum, Hebert, and Ikeuchi proposed a 
method that has $O(n^2)$ time complexity \cite{Shu}. However, it is not very practical 
for the purpose of data retrieval. 
Another disadvantage of the continuous curvature method is that 
the calculation of curvatures return real numbers and that introduces 
precision errors. \cite{Mok} \cite{Heb} \cite{Hil} 
A multi-scaling method was proposed to complete this task, but it takes more time.    

The technique presented in this section is not to attempt to replace the methods developed
before. We try to explore more applications using digital curvatures. As we discussed 
in Section 2 and 3, the local digital curvature is determined by the 
local shape of the digital surface. The number of each type of surface-point may indicate 
the features of a 3D object.

\subsection {Feature Vectors of 3D object Based on Digital Curvatures} 

First, form a feature vector that only contains the number of digital surface points
in each of the categories $M_3$, $M_4$, $M_5$, and $M_6$.

A 3D digital object may not necessarily be a 3D digital manifold. This is because
we have strict definitions for 3D manifolds.  
However, this does not affect the use of curvatures.

The feature vector is
\begin{equation}
 fv =\frac{1}{T} (|M_3|,|M_4|,|M_5|,|M_6|) = (r_3,r_4,r_5,r_6) 
\end{equation} 

\noindent where $T$ is the total surface points.

For example, an object has a total of 1679 digital points. 
$|M_3|=469;  |M_4|=995;  |M_5|= 183; |M_6|= 32$. Because of some non-manifold points, we  
corrected the data for |M|'s: 484;  995;  180;  28 ; 240. It also includes the total of 8
none manifold points.  Therefore, we get the 
Ratio $r_3=0.288267$, $ r_4= 0.592615$, $r_5 = 0.107207$, and $r_6=  0.016677$.

The Euclidean distance of two feature vectors:

\[d=\sqrt {\Sigma_{i=3}^{6}((xi-yi)^2)}\] 

We can also calculate other distances such as the general Minkowski distance.

\[d_{n}= ( \Sigma_{i=3}^{6}((x_i-y_i)^n))^{1/n} \]

We could also use the scaling method from Section 3 to get more features. Here we 
use this method to do a rough classification for 3D shapes. 
 
The computing examples are provided in Princeton Benchmark Website. \cite{Fun}  

\begin{figure}[h]
	\begin{center}

   \epsfxsize=3in 
   \epsfbox{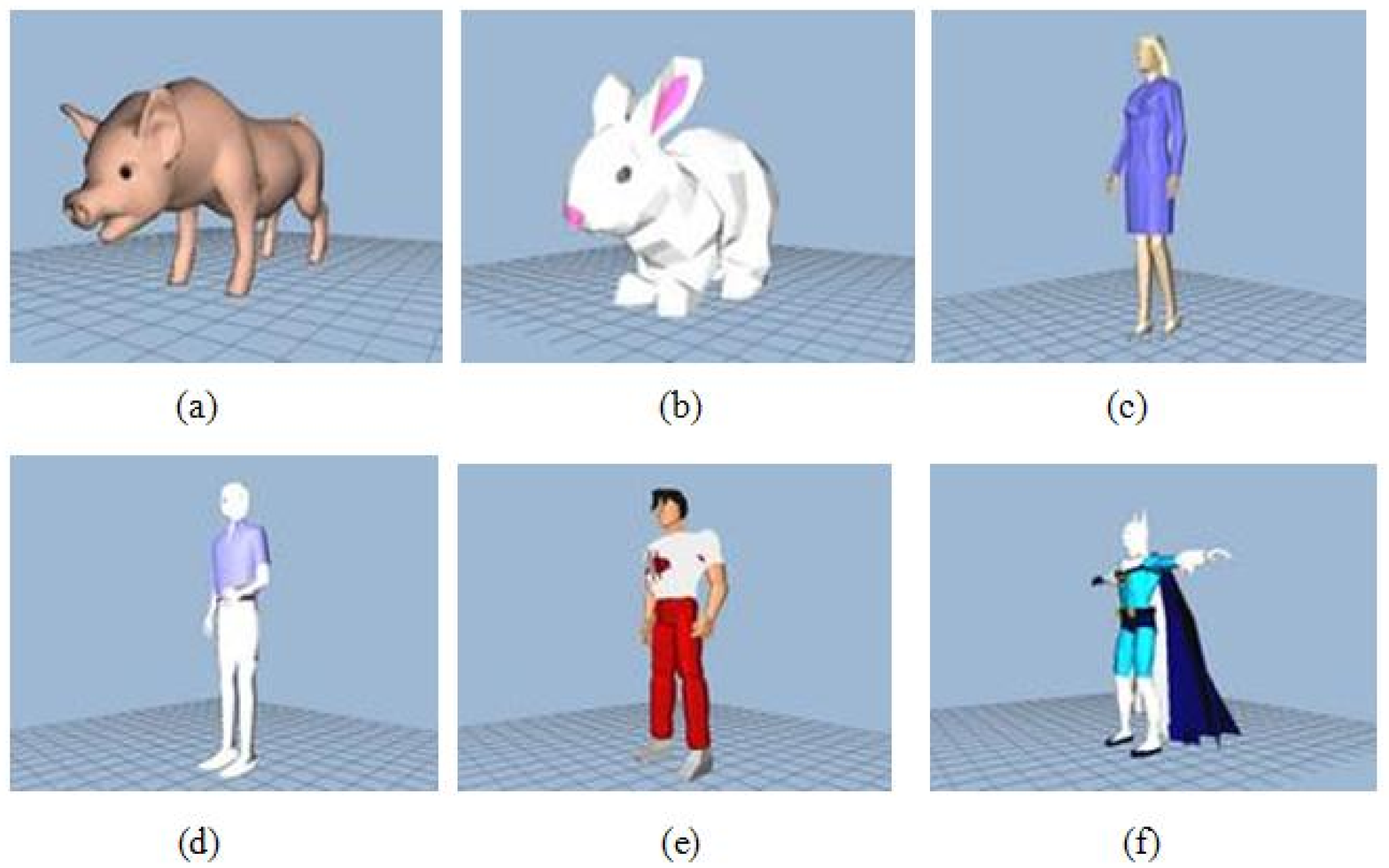}

	\end{center}
\caption{ Six 3D objects from Princeton database}
\end{figure}

\subsection {Similarity and Distance Analysis based on the Feature Vectors }

Feature ratio vectors $ex= (r3,r4,r5,r6)$ where $r_k = |M_k|/T$,

 $e_1= (0.288267,  0.592615,  0.107207,  0.016677) $
 
 $e_2= (0.262424,  0.508752,  0.193369,  0.044133)$ 
 
 $e_3 = (0.168149,  0.680220,  0.144854,  0.008895)$ 
 
 $e_4 = (0.152833,  0.711492,  0.122506,  0.013966)$

 $e_5 = (0.148500,  0.710425,  0.135432,  0.007128)$
 
 $e_6 = (0.162700,  0.688310,  0.140705,  0.010093)$

The Distance matrix of two-vector pairs is

\begin{eqnarray}
    \left (\begin{array}{llllll}
    0            &               &              &             &    &           \\
    0.015878586  &   0           &              &             &    &           \\ 
    0.023580826  &  0.041884473  &  0           &             &   &            \\
    0.032715518  &  0.059045308  &  0.001737666 &  0           &   &           \\
    0.034301844  &  0.058376743  &  0.001390322 &  0.000233753 &    0   &       \\
    0.02609007   &  0.04611817   &  0.000113789 &  0.000980967 &   0.000727309 & 0

              \end{array}
              \right )
\end{eqnarray}

In the matrix, $a_{ij}$ describes the distance between object $i$ and object $j$.
 The matrix is symmetric so we only present half of it.   
 We can easily see that Objects 1 and Object 2 are closed related. Objects 4,5, and 6
 are similar. From the data pictures displayed in Fig. 10, we can see that is correct. 
An interesting observation is that Object 3 does not go with three other objects in the 
second category.    \\\\

{\it Acknowledgement} The authors would like to thank
Professor Rama Chellappa for discussion, putting ideas, and working jointly in this 
project. We also like to thank NIST Face Recognition Grand Challenge (FRGC) and Princeton 
3D datasets.












\end{document}